\documentclass[twocolumn,
superscriptaddress,
amsmath,amssymb,aps,prl]{revtex4-1}

\usepackage{graphicx}
\usepackage{dcolumn}
\usepackage{bm}
\usepackage{multirow}
\usepackage{array}
\usepackage{diagbox}

\usepackage[colorlinks,urlcolor=blue,citecolor=blue,linkcolor=blue]{hyperref}

\usepackage{comment}
\usepackage{color}
\usepackage{physics}

\renewcommand{\k}{{\bf k}}
\newcommand{\p}{{\bf p}}

\newcommand{\q}{{\bf q}}
\newcommand{\Q}{{\bf Q}}

\newcommand{\0}{{\bf 0}}

\newcommand{\nn}{\nonumber}
\newcommand{\beq}{\begin{equation}}
\newcommand{\eeq}{\end{equation}}

\newcommand{\vac}{\ket{0}}

\newcommand{\FS}{\ket{\mathrm{FS}}}
\newcommand{\lFS}{\bra{\mathrm{FS}}}

\newcommand{\vkq}{v_{\k\q}}
\newcommand{\abf}{a_\mathrm{BF}}
\newcommand{\abb}{a_\mathrm{BB}}
\newcommand{\kf}{k_\textrm{F}}

\begin{document}

\title{Quantum droplets in a resonant Bose-Fermi mixture}

\author{Sam Foster}
\affiliation{School of Physics and Astronomy, Monash University, Victoria 3800, Australia}

\author{Olivier Bleu}
\affiliation{School of Physics and Astronomy, Monash University, Victoria 3800, Australia}
\affiliation{Institut f{\"u}r Theoretische Physik, Universit{\"a}t Heidelberg, 69120 Heidelberg, Germany}

\author{Jesper Levinsen}
\affiliation{School of Physics and Astronomy, Monash University, Victoria 3800, Australia}

\author{Meera M. Parish}
\affiliation{School of Physics and Astronomy, Monash University, Victoria 3800, Australia}

\date{\today}

\begin{abstract}
We study the canonical problem of a Fermi gas interacting with a weakly repulsive Bose-Einstein condensate at zero temperature. To explore the quantum phases across the full range of boson-fermion interactions, we construct a versatile variational ansatz that incorporates pair correlations and correctly captures the different polaron limits. Remarkably, we find that self-bound quantum droplets can exist in the strongly interacting regime, preempting the formation of boson-fermion dimers, when the Fermi pressure is balanced by the resonant boson-fermion attraction. This scenario can be achieved in experimentally available Bose-Fermi mixtures for a range of boson-fermion mass ratios in the vicinity of equal masses. 
We furthermore show that a larger fermion density instead yields phase separation between a Bose-Fermi mixture and excess fermions, as well as behavior reminiscent of a liquid-gas critical point. Our results suggest that first-order quantum phase transitions play a crucial role in the phase diagram of Bose-Fermi mixtures. 
\end{abstract}

\maketitle

Quantum mixtures of bosons and fermions are fundamental to a range of different systems in physics. 
Beginning with liquid $^4$He-$^3$He mixtures, which were studied as early as the 1960's~\cite{London1962,Edwards1965,Anderson1966-HeatCapacity,Anderson1966-Magnetic, Bardeen1967}, Bose-Fermi mixtures have since been realized in ultracold atomic vapors~\cite{Schreck2001-1,Ferrari2002,Roati2002,Goldwin2004,Inouye2004,Ospelkaus2006,%
Hu2016,DeSalvo2017,Lous2018,DeSalvo2019,Yan2020,Fritsche2021,Chen2022,Duda2023,Patel2023,Baroni2024,Yan2024}, and even in the optical response of doped semiconductors~\cite{Sidler2017}, where bosonic excitons %
are optically introduced into a gas of fermionic charge carriers such as electrons. In particular, the tunability of the cold-atom system has enabled Bose-Fermi mixtures to be investigated for a large variety of boson-fermion mass ratios and interactions. However, the mixed statistics make the many-body phase diagram  exceptionally challenging to describe beyond the weak-coupling regime~\cite{Viverit2000,Roth2002,Viverit2002,Albus2002,DukelskyComment}, and there remain outstanding questions regarding the nature of the quantum phase transitions that are associated with 
the quantum depletion of the Bose-Einstein condensate (BEC)~\cite{Powell2005,Marchetti2008,Bertaina2013,Duda2023}.

\begin{figure}
    \centering
    \includegraphics[width=0.9\linewidth]{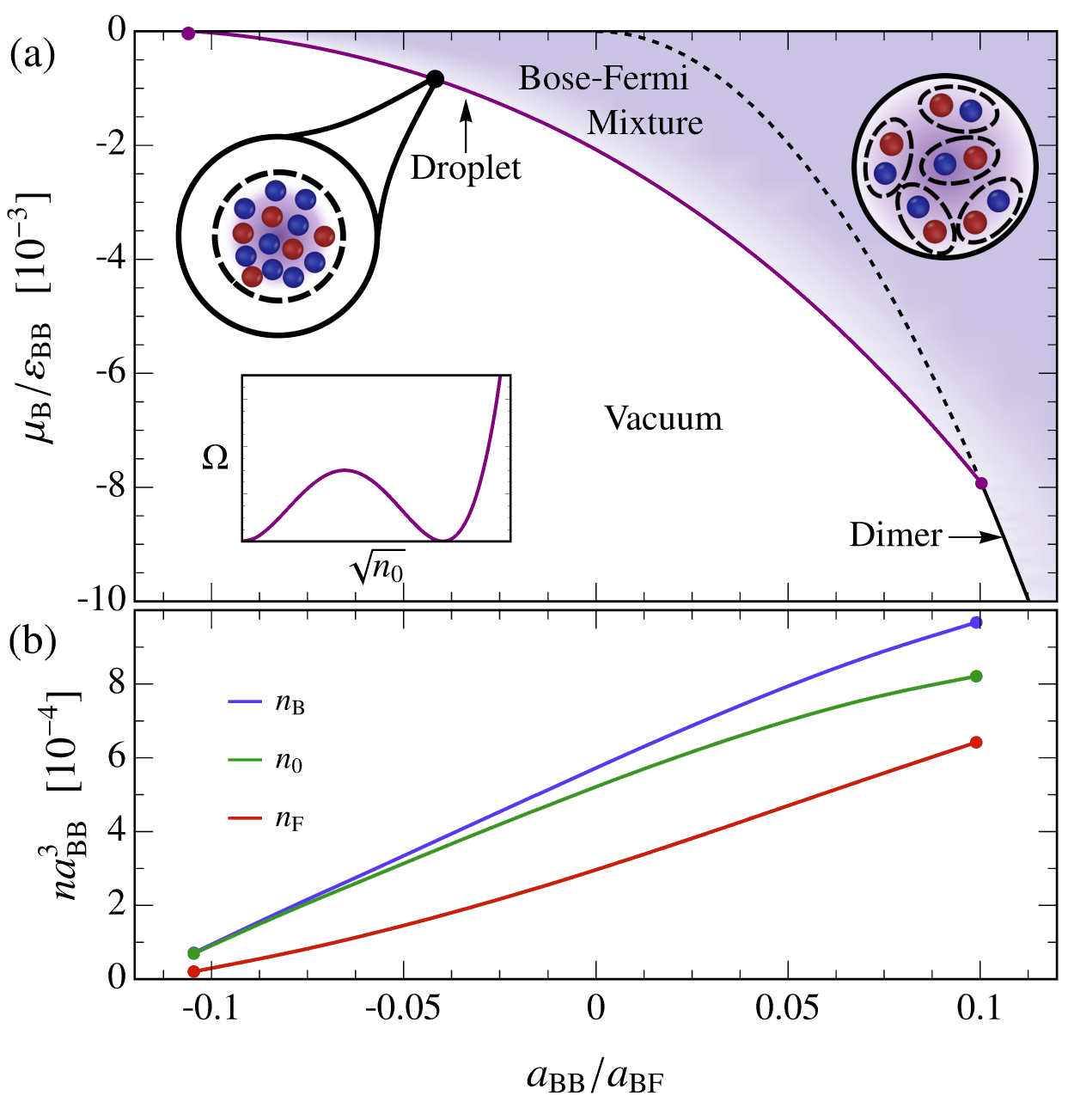}
    \caption{(a) Phase diagram of a Bose-Fermi mixture as a function of $\mu_\mathrm{B}$ and interaction $a_\mathrm{BF}$ at fixed $\mu_\mathrm{F}=0$. Here we take $m_\mathrm{B}/m_\mathrm{F}=23/40$, and we scale lengths by $a_\mathrm{BB}$ and energies by $\varepsilon_{\mathrm{BB}}= 1/m_\mathrm{B} a_{\mathrm{BB}}^2$.  The quantum droplet exists along the first-order phase boundary separating a uniform BF mixture~(purple) from the vacuum~(white), and it corresponds 
    to a minimum in the free energy which is degenerate with the vacuum (inset).
    Beyond the purple circles, the droplet is replaced by a second-order phase boundary, either between vacuum and an atomic Bose gas at $\mu_\mathrm{B}=0$, or between vacuum and a Fermi gas of dimers where $\mu_\mathrm{B}$ corresponds to the dimer energy $-1/2m_\mathrm{r} a_\mathrm{BF}^2$ (black line). 
    (b) Boson, condensate, and fermion densities~(blue, green, red) %
    in the droplet phase.}
    \label{fig:mubAbfDroplet}
\end{figure}

Here we show that quantum-droplet phases and liquid-gas critical behavior emerge for resonant boson-fermion attraction (Fig.~\ref{fig:mubAbfDroplet}). First predicted for %
Bose-Bose (BB) mixtures~\cite{Petrov2015}, a quantum droplet is a novel self-bound, liquid-like cluster of particles that directly results from the balance between mean-field attraction and repulsive beyond-mean-field %
quantum fluctuations~\cite{LHY1957}. Such droplets have already been realized experimentally in bosonic $^{39}$K spin mixtures~\cite{Cabrera2018,Semeghini2018} and $^{41}$K-$^{87}$Rb mixtures~\cite{Derrico2019,Burchianti2020}, as well as in dipolar Bose gases~\cite{Schmitt2016,FerrierBarbut2016, Chomaz2016}.
However, while quantum droplets have been predicted to occur in Bose-Fermi (BF) mixtures~\cite{Rakshit2019SciPost,Rakshit2019}, they have not yet been observed. 
A key difficulty is that current descriptions of BF droplets are based on second-order perturbation theory~\cite{Viverit2002} and are thus restricted to the weak-coupling limit, similarly to the case of BB droplets~\cite{Petrov2015}. Yet the predicted BF droplets lie beyond the weak-coupling regime~\cite{Rakshit2019SciPost,Cui2018}. Moreover, recent experiments~\cite{Patel2023} show that lowest-order mean-field theory is insufficient to capture the instability of BF mixtures with increasing attraction, and one must instead use strong-coupling theories that exactly describe the boson-fermion two-body scattering~\cite{Fratini2010,Fratini2012,Guidini2014,Guidini2015,Shen2024,Gualerzi2025,Pisani2025}.

In this Letter, we introduce a new variational ansatz that describes a Bose-Fermi mixture across the full range of boson-fermion interactions 
at zero temperature. %
Importantly, it recovers perturbation theory in the weak-coupling limit as well as 
correctly capturing the two extreme limits of population imbalance, corresponding to Bose and Fermi polarons~\cite{Massignan2014, Scazza2022,Grusdt2025}, where 
a single impurity is dressed by excitations of a Bose or Fermi gas, respectively. 
Our work also goes beyond the variational approach for BF mixtures in Ref.~\cite{Yu2011} which was restricted to the limit of a small density of Fermi polarons.
By minimizing the free energy obtained from our ansatz, 
we find that quantum droplets can exist for resonant boson-fermion attraction and beyond, preempting the formation of bound boson-fermion dimers. 
We furthermore reveal the existence of liquid-gas critical behavior where, for instance, 
a BF liquid is surrounded by an atomic Fermi gas.
Our results are important for cold-atom experiments since stable BF mixtures have now been achieved in the strong-coupling regime~\cite{Yan2020,Fritsche2021,Chen2022,Patel2023,Duda2023,Baroni2024,Yan2024}, and we expect our droplet phases to be accessible for a range of mass ratios around equal masses.

{\it Model}---We model the three-dimensional Bose-Fermi mixture using a Hamiltonian which accounts for both BF and BB interactions:
\begin{multline}\label{eq:complete-hamiltonian}
    \mathcal{H} = 
    \sum_{\k} \bigg [
    (\epsilon_{{\k}}^\mathrm{F}-\mu_\mathrm{F}) f_{\k}^\dagger f_{\k}
    + (\epsilon_{\k}^\mathrm{B}-\mu_\mathrm{B}) b_{\k}^\dagger b_{\k}
    \bigg ] \\ 
    + g_\mathrm{BF} 
    \sum_{\k \k' \q} f^\dagger_{\k + \q} 
    b^\dagger_{\k' - \q}   b_{\k'}f_{\k}
    + \frac{g_\mathrm{BB} }{2}
    \sum_{\k \k' \q} b^\dagger_{\k + \q} 
    b^\dagger_{\k' - \q}   b_{\k'}b_{\k},
\end{multline} 
where $f_\k^\dag\ (b_\k^\dag)$ creates a fermion~(boson) with momentum $\k$. The corresponding dispersions are $\epsilon_{\k}^\mathrm{F/B} = |\k|^2/2m_\mathrm{{F/B}} \equiv k^2/2m_\mathrm{{F/B}}$ with masses $m_\mathrm{F/B}$ and chemical potentials $\mu_\mathrm{F/B}$. 
We work in units such that $\hbar$ and the system volume are set to 1. 

The atoms interact via short-range contact potentials. In order to describe a droplet phase in the regime of strong boson-fermion attraction, we consider their scattering to arbitrary order which leads to the renormalization condition
$\frac{1}{g_\mathrm{BF}} 
    =
    \frac{m_\mathrm{r}}{2\pi a_\mathrm{BF}} 
    -
    \sum_{\k}^{\Lambda}
    \frac{1}{\epsilon_{\k}^\mathrm{F} + \epsilon_{\k}^\mathrm{B}}.$
Here, $m_\mathrm{r} = m_\mathrm{F} m_\mathrm{B}/(m_\mathrm{F} + m_\mathrm{B})$ is the reduced mass, and $\Lambda$ is an ultraviolet cutoff, which will be sent to infinity at the end of the calculation. 
This procedure allows us to replace the model parameters $g_\mathrm{BF}$ and $\Lambda$ by the physical parameter, the BF scattering length $a_\mathrm{BF}$, with our model featuring a two-body bound state when $a_\mathrm{BF}>0$. On the other hand, the boson-boson interaction is taken to be weakly repulsive and treated at the mean-field level such that $g_\mathrm{BB}= 4\pi a_{\mathrm{BB}}/m_\mathrm{B}$, with scattering length $a_{\mathrm{BB}}>0$.

\textit{Variational ansatz---}To describe the ground state of the interacting BF mixture, we draw inspiration from 
the variational wave function based on a coherent state of fermion pairs~\cite{Schrieffer1964}, which has successfully been used to investigate the BCS-BEC crossover in strongly interacting two-component Fermi gases~\cite{Randeria2012}. 
Specifically, we introduce the ansatz
\begin{equation}\label{eq:ansatz-exp}
    \ket{\psi} = \mathcal{N}e^{\lambda\beta^\dag}\FS=\mathcal{N}e^{\alpha_0 b_0^\dag+\sum_{\k\q} \alpha_{\k\q} b^\dag_{\q-\k} f^\dag_\k f_\q} \FS,
\end{equation}
where the Fermi sea $\FS = \prod_{q \leq k_\mathrm{F}} f_\q^\dag \vac$ with density $n_\mathrm{F}$ is filled up to the Fermi momentum $k_\mathrm{F} =(6\pi^2 n_\mathrm{F})^{1/3}$, and $\vac$ is the vacuum. The operator $\beta^\dag$ effectively creates a quasiparticle involving a boson dressed by particle-hole fluctuations of the Fermi sea, where $\lambda$ is the corresponding amplitude which is absorbed into the variational coefficients $\alpha_0$ and $\alpha_{\k\q}$. 
Here, and in the following, we assume that $q \leq k_\mathrm{F} < k$, which ensures that the operator $B_{\k\q}^\dag \equiv b^\dag_{\q-\k} f^\dag_\k f_\q$ satisfies the relation $[B^\dag_{\k'\q'},B^\dag_{\k\q}] = 0$, thus allowing one to freely manipulate the terms in the exponential in Eq.~\eqref{eq:ansatz-exp}. 
Additionally,  $\mathcal{N}$ is a normalization constant (which we can take to be real), and the condensate density is $n_0 = |\alpha_0|^2$. 

By %
introducing variables $u_\q$ and $v_{\k\q}\equiv u_\q \alpha_{\k\q}$, we find that Eq.~\eqref{eq:ansatz-exp} %
can be written as
\begin{equation}\label{eq:ansatz-bcs-form}
\hspace{-2mm}    \ket{\psi} = e^{-|\alpha_0|^2/2} e^{\alpha_0 b_0^\dag}
    \prod_{\q} (u_{\q} + \sum_{\k} v_{\k \q} b^\dag_{\q-\k} f^\dag_\k f_\q)\FS.
\end{equation}
Our coherent state now strongly resembles the celebrated BCS wave function~\cite{Schrieffer1964}. 
Equation~\eqref{eq:ansatz-bcs-form} represents a remarkably flexible ansatz, since 
it can describe the limit of a non-interacting BF mixture (where $v_{\k\q} \to 0$) as well as a Fermi gas of tightly bound boson-fermion pairs (where $u_\q \to 0$). %
In addition, our ansatz is not limited to fixed population ratios unlike other coherent-state ans\"{a}tze such as that recently proposed to describe a quartet superfluid~\cite{Liu2023QuartetSuperfluid}.
It can therefore straightforwardly describe the Fermi and Bose polaron problems of extreme population imbalance~\cite{supp}, reducing in both cases to the highly successful ansatz of including at most a single excitation of the medium~\cite{Chevy2006,Li2014BosePolaron}. %
However, a unique feature of our ansatz is that it also enables us to extract the effective interactions between these quasiparticles~\cite{inPrep}. Here, we focus on the tantalizing prospect of realizing BF droplets in a resonantly interacting mixture.

To proceed, we must deal with the non-trivial correlations arising from the multiple momenta in the composite operator $\beta^\dag$, which make determining the normalization and other expectation values a challenge. These correlations are present in the commutator:
\begin{align} \label{eq:commutator}
       \nn  [B_{\k'\q'},B^\dag_{\k\q}]
    = & \,
    \delta_{\q\q'} \delta_{\k\k'} (1-f^\dag_{\k} f_{\k} + b_{\q-\k}^\dag b_{\q-\k})
    \\
    &-\delta_{\q\q'} f^\dag_{\k} f_{\k'} b^\dag_{\q-\k} b_{\q-\k'}
    + f_{\q} f^\dag_{\q'}(\cdots).
\end{align} 
This would reduce to a simple boson commutation relation if we could neglect the additional $f$ and $b$ operators appearing on the right hand side.  As we show in the Supplemental Material~\cite{supp}, we can choose an arrangement of the operators in the ansatz such that the final term in Eq.~\eqref{eq:commutator}, i.e., everything involving $f_{\q} f^\dag_{\q'}$, vanishes when we evaluate the normalization. This then leaves terms that are also expected to be small: the number operators in the first line approximately cancel since the fermion and boson distributions coincide in the limit $k \gg q$, while the remaining four-operator term in the second line leads to higher order correlations involving multiple fermions, which is suppressed due to Pauli exclusion.  Thus, neglecting these small terms, we arrive at the normalization $\mathcal{N} \simeq e^{- |\alpha_0|^2/2}/{\prod_{\q} \sqrt{1 + \sum_{\k} |\alpha_{\k \q}|^2}},$ which simply implies that $u_\q^2 + \sum_{\k} |v_{\k\q}|^2 =1$~\cite{supp}. In the following, we can take the parameters to be real without loss of generality.

\textit{Free energy---}The phase diagram of the mixture is governed by the free energy, $\Omega\equiv \bra{\psi}\mathcal{H}\ket{\psi}$.  %
Within our ansatz,  it %
takes the form~\cite{supp}
\begin{align} \nn
    \Omega
    = &
    \sum_{\q} ( \epsilon_{\q}^\mathrm{F} - \mu_\mathrm{F} )
    -\mu_\mathrm{B} n_0
    +
    \frac{g_\mathrm{BB}}{2} n_0^2
    +\sum_{\k \q} E_{\k \q}v_{\k \q}^2
    \\&+
    2g_\mathrm{BF}\sqrt{n_0}\sum_{\k \q} u_{\q} v_{\k \q}     +g_\mathrm{BF} \sum_{\k \k' \q} v_{\k \q} v_{\k' \q}
    , \label{eq:freeenergy}
    \end{align}
where we define ${E}_{\k\q} \equiv\epsilon_{\k}^\mathrm{F}-\epsilon_{\q}^\mathrm{F}+\epsilon_{\q-\k}^\mathrm{B}-\mu_\mathrm{B} + 2 g_\mathrm{BB} n_0$. We also have the total boson density $n_\mathrm{B} = n_0 + \sum_{\k \q} v_{\k\q}^2$, which %
clearly shows that $v_{\k\q}$ is associated with quantum fluctuations.

To obtain the free energy $\Omega(\mu_\mathrm{F},\mu_\mathrm{B})$ in the ground state, we must minimize with respect to the variational parameters $v_{\k\q}$, the condensate order parameter $\alpha_0$, and the fermion density $n_\mathrm{F}$. %
By first considering the minimization associated with quantum fluctuations, %
\begin{equation} \label{eq:min-akq}
    E_{\k \q} v_{\k \q}=
    (\sqrt{n_0} v_{\k \q}-u_{\q})
    g_{\mathrm{BF}} \sum_{\k'} \frac{v_{\k' \q}}{u_\q}
    -
    g_{\mathrm{BF}}\sqrt{n_0} u_{\q},
\end{equation}
we can define the renormalized quantity
\begin{align}\label{eq:chi}
    \chi_{\q} \equiv
    \frac{g_{\mathrm{BF}}}{\sqrt{n_0}} \sum_{\k} \frac{v_{\k\q}}{u_\q}
    =
    \Bigg[\frac{1}{g_{\mathrm{BF}}}
    +
    \sum_{\substack{\k}}
    \frac{1}{E_{\k\q}   - n_0 \chi_\q}
    \Bigg]^{-1}.
\end{align}
This is akin to a self-consistent in-medium $T$-matrix, an approach that has, so far, only been considered for a BF mixture in the normal state at finite temperature~\cite{Manabe2021}. 

Using Eqs.~\eqref{eq:min-akq} and \eqref{eq:chi}, we can eliminate $v_{\k\q}$ in the free energy, yielding the more physically transparent form
\begin{align}
       \Omega =&
    \frac{3}{5} E_\mathrm{F} n_\mathrm{F}-\mu_\mathrm{F}n_\mathrm{F}
    -\mu_\mathrm{B}n_0 
    + \frac{g_{\mathrm{BB}} }{2} n_0^2 +n_0 \sum_{\q} \chi_{\q} , 
    \label{eq:free-energy-renormalised}
\end{align}
where we have defined the Fermi energy $E_\mathrm{F}=k_\mathrm{F}^2/2 m_\mathrm{F}$. %
The remaining equations of motion, $\partial \Omega/\partial n_0 = 0$ and $\partial \Omega / \partial n_\mathrm{F}=0$, now take the form %
\begin{subequations}
    \label{eq:min-conditions}
    \begin{align}
         &\mu_\mathrm{B} = g_{\mathrm{BB}} (2n_\mathrm{B} -n_0)+
    \sum_{\q} u_{\q}^2 \chi_\q 
    ,
    \label{eq:min-n0}
    \\
        &\mu_\mathrm{F} = 
    E_\mathrm{F} + n_0 \chi_{k_\mathrm{F}}
    + \sum_{\q} (E_{k_\mathrm{F} \q} - n_0 \chi_{\q})
    v_{k_\mathrm{F} \q }^2.
    \label{eq:min-kf}
    \end{align}
\end{subequations}
In the limit of weak BF interactions, 
where $n_\mathrm{B} \to n_0$ and $\chi_\q\to 2 \pi a_\mathrm{BF}/m_\mathrm{r}$,
we find that Eq.~\eqref{eq:free-energy-renormalised} 
reduces to the expected mean-field expression. Moreover, expanding to second order in $a_\mathrm{BF}$ recovers the perturbative result from Ref.~\cite{Viverit2002} for the energy density $\mathcal{E} \equiv \Omega + \mu_\mathrm{B} n_\mathrm{B} + \mu_\mathrm{F} n_\mathrm{F}$~\cite{supp}.

\textit{Bose-Fermi Droplets---}The quantum-droplet phase corresponds to the coexistence between vacuum and a BF mixture at finite density~\cite{Petrov2025varenna}. 
This immediately implies that both configurations are minima of the free energy with vanishing pressure $P=-\Omega=0$ [see inset in Fig.~\ref{fig:mubAbfDroplet}(a)]. 
To ensure that the free energy has a minimum at the vacuum phase where $\alpha_0 =0$, we require 
$\mu_\mathrm{B}\leq 0$ like for the case of droplets in binary Bose mixtures~\cite{Petrov2015}.
However, since the Fermi pressure of an ideal Fermi gas (i.e., the Bose vacuum) scales as $P\propto\mu_\mathrm{F}^{5/2}$, we must have $\mu_\mathrm{F}=0$.

\begin{figure}
    \centering
    \includegraphics[width=.88\linewidth]{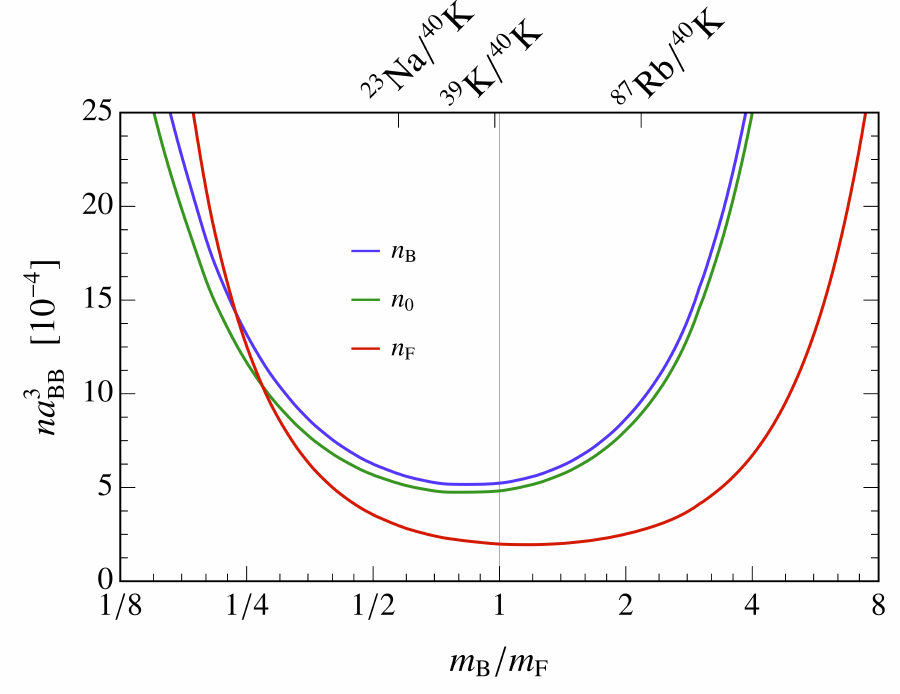}
    \caption{Densities in the quantum droplet phase at unitarity, $%
    1/a_\mathrm{BF} = 0$. The boson and fermion densities are lowest for %
    mass ratios in the vicinity of equal masses, which can be accessed with experimentally available BF mixtures (indicated on the top axis). 
   }
    \label{fig:mubMassRatioDroplet}
\end{figure}

To investigate the existence of droplets, we thus plot the phase diagram as a function of $a_\mathrm{BB}/a_\mathrm{BF}$ and $\mu_\mathrm{B} <0$ at fixed $\mu_\mathrm{F}=0$, as shown in Fig.~\ref{fig:mubAbfDroplet}(a). 
Here, we %
show our results for the experimentally relevant Na-K mixture with $m_\mathrm{B}/m_\mathrm{F} =23/40$, which lies in the optimal range of mass ratios for droplets (see Fig.~\ref{fig:mubMassRatioDroplet}), as we discuss further below. %
Crucially, we find that the droplet phase can exist 
when the ratio of the two scattering lengths satisfies 
$-0.1\lesssim a_\mathrm{BB}/a_\mathrm{BF}\lesssim0.1$, and that it occurs along the phase boundary that represents a first-order phase transition between a uniform BF mixture and the vacuum phase. 
Of particular interest is the existence of a BF droplet at unitarity, $1/a_\mathrm{BF}=0$, in contrast to other droplet scenarios which typically require weak interactions. 
This behavior is connected to the fact that the scaling of the energy with density $n$
is the same for the Fermi pressure, $\sim n^{5/3}$, as it is for the BF attraction, $\sim \abf n_\mathrm{B}n_\mathrm{F}\sim \abf n^2$, where $\abf$ is replaced by $\sim 1/n^{1/3}$ at unitarity. 
We can therefore understand the existence of this phase at unitarity as being due to the delicate balance between BF attraction, BB repulsion, and Fermi pressure. 
The end of the droplet phase at $a_\mathrm{BF}>0$ corresponds to when pairing becomes energetically favorable, yielding a second-order transition from vacuum to a Fermi gas of BF dimers.

To gain further insight into the BF droplet, Fig.~\ref{fig:mubAbfDroplet}(b) shows the associated Bose and Fermi densities. We see that the droplet involves a greater density of bosons and it remains very dilute even close to unitarity, where the quantum depletion of the Bose gas remains small $\lesssim 15\%$. Furthermore, the gas parameter $n_\textrm{B}\abb^3$ at unitarity is an order of magnitude smaller than that considered in quantum Monte Carlo calculations of BF mixtures~\cite{Bertaina2013,Guidini2015}.
We note that the critical interaction $a_\mathrm{BF}/a_\mathrm{BB}$ for the onset of droplet formation depends on the mass ratio~\cite{supp} and we find values that are similar to that expected from second-order perturbation theory~\cite{Rakshit2019}. %
However, our model predicts noticeably larger densities, indicating that droplets exist outside the strictly perturbative regime, and indeed we find that the onset value is $k_\mathrm{F} a_\mathrm{BF}\approx -1$ for a range of mass ratios. 

While our predicted unitary BF droplets potentially extend to all mass ratios, the boson and fermion densities within the droplet can vary significantly, as highlighted in Fig.~\ref{fig:mubMassRatioDroplet}. In particular, we see that the densities diverge when $m_\textrm{B} \to \infty$ or $m_\textrm{F} \to \infty$, since the boson-boson repulsion or the Fermi pressure respectively vanish in these limits. Indeed, the suppressed Fermi pressure in the case of very heavy fermions leads to a greater density of fermions than bosons within the droplet.
Therefore, BF mixtures with mass ratios in the vicinity of equal masses are the optimal for observing %
droplets since they yield the lowest densities. 
For instance, a Na-K mixture with a typical atomic BEC density of $n_\mathrm{B}=10^{15}\,\mathrm{cm}^{-3}$~\cite{PethickSmithBook} and K density of $n_\mathrm{F}=5\times 10^{14}\,\mathrm{cm}^{-3}$, we find $a_\mathrm{BB}=80a_0$ at the onset of the droplet phase ($a_0$ is the Bohr radius), while $a_\mathrm{BB}=160a_0$ for a droplet at the BF resonance. These densities and scattering lengths are comparable with typical experiments.

In practice, droplet formation competes with losses due to three-body recombination and with the formation of few-body states such as Efimov trimers~\cite{Efimov1971} (2 bosons and 1 fermion) or Kartavtsev-Malykh (KM) trimers~\cite{Kartavtsev2007} (2 fermions and 1 boson). However, we stress that three-body losses and Efimov trimers are both strongly suppressed when $m_\mathrm{B}\sim m_\mathrm{F}$. Indeed, recent experiments on Na-K Bose-Fermi mixtures have successfully realized thermalized Bose polarons at unitarity~\cite{Yan2020} and a resonant mixture without collapse~\cite{Duda2023}, thus demonstrating the feasibility of strongly correlated BF mixtures with moderate mass ratios. On the other hand, if $m_\mathrm{B}/m_\mathrm{F}\gg 1$ then we expect Efimovian three-body correlations to dominate~\cite{Braaten2006}, while in the opposite limit where $m_\mathrm{F}/m_\mathrm{B}\gtrsim8.2$ we have KM trimers~\cite{Kartavtsev2007}. %

\begin{figure}
    \centering
    \includegraphics[width=1\linewidth]{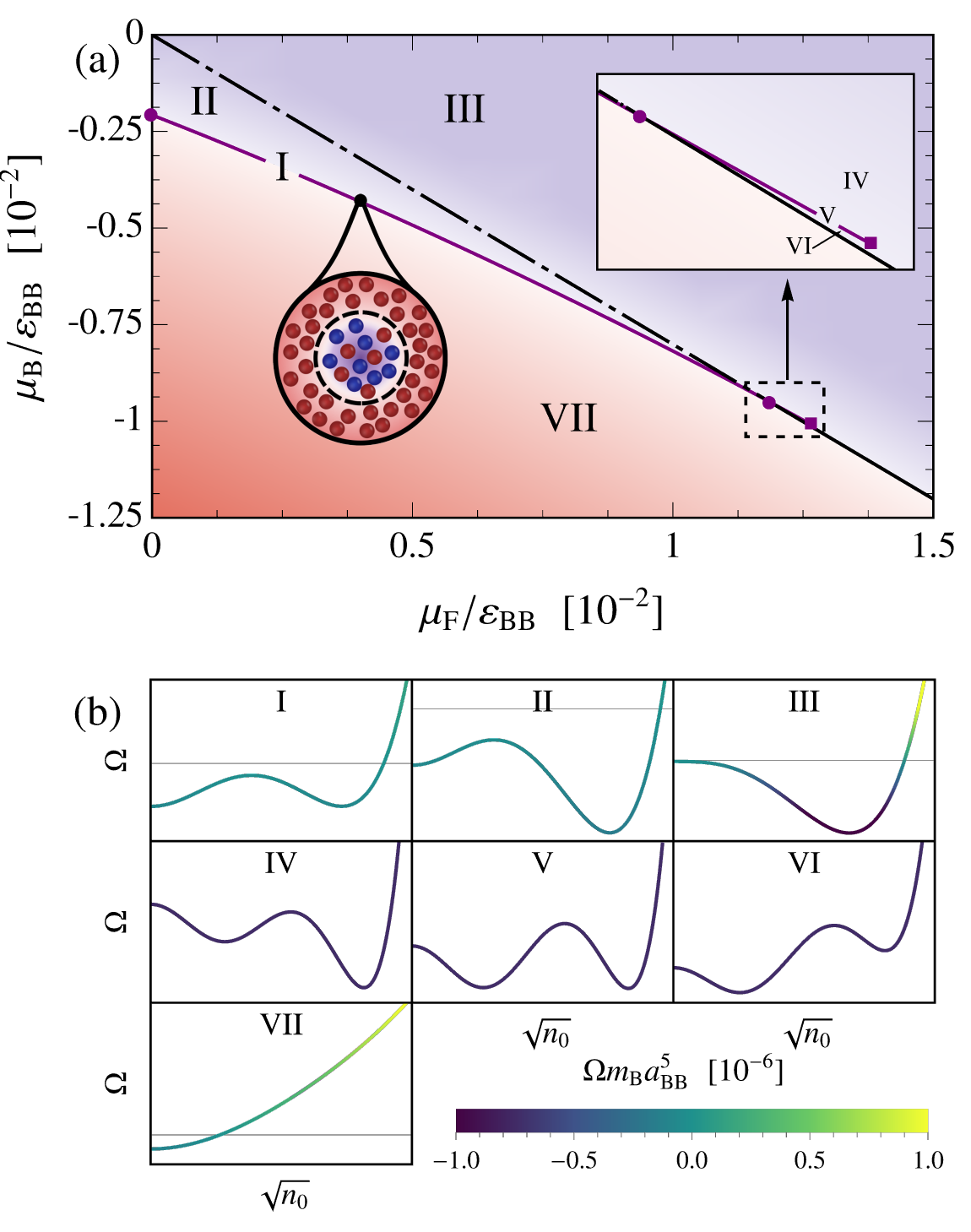}
    \caption{ 
    (a) Phase diagram of a Bose-Fermi mixture at unitarity %
    for $m_\mathrm{B}/m_\mathrm{F}=23/40$. %
    The red and purple regions denote an ideal Fermi gas and BF mixture, respectively. 
    The purple line corresponds to a first-order phase boundary that evolves from a BF droplet at $\mu_\mathrm{F}=0$ to liquid-gas phase separation (lines I and V) before terminating in a critical end point (square, zoomed-in inset), where two finite-density minima in the free energy merge. 
The solid black line denotes a second-order phase boundary where $\mu_\mathrm{B}$ corresponds to the polaron energy of a single bosonic impurity~\cite{supp}. 
    Following its intersection~(circle) with the purple line, it becomes a spinodal line (dot-dashed) separating regions \rm{II} and \rm{III}. 
    (b)  The corresponding free energies for the regions labeled by roman numerals in (a). 
    Each panel has a different scale, with the value of the free energy indicated by the color bar. %
    } 
    \label{fig:mubmufDroplet}
\end{figure}

\textit{Quantum liquid-gas transitions---}Finally, we investigate the evolution of the droplet as we increase the fermion chemical potential $\mu_\mathrm{F}$ from zero,
as shown in Fig.~\ref{fig:mubmufDroplet}. 
In this case, the pressure no longer vanishes and we find that for certain parameters (line marked I) the excess fermions are expelled from the droplet such that 
we instead have a BF liquid surrounded by 
a coexisting ideal Fermi gas. This liquid-gas phase separation corresponds to two degenerate minima in the free energy which define the first-order phase boundary between a Fermi gas (i.e., the boson vacuum) and a uniform BF mixture.

For large fermion densities (i.e., large $\mu_\mathrm{F}$), we find that the boson density in the BF mixture can continuously go to zero as $\mu_\mathrm{B}$ is lowered, thus indicating a change to a second-order transition between a BF mixture and the Fermi gas. 
We can understand this transition by evaluating the Bose chemical potential in the limit of an arbitrarily small admixture of bosons immersed in an ideal Fermi gas. This limit precisely constitutes the Fermi-polaron problem~\cite{Massignan2014,Scazza2022}, and indeed we find that the phase boundary now follows the Fermi-polaron energy calculated within the celebrated Chevy ansatz~\cite{Chevy2006,supp}.

As the fermion density is lowered, the continuous transition is eventually preempted by the first-order liquid-gas transition, such that 
the polaron energy instead defines a spinodal line where the metastable minimum at small $\alpha_0$ is lost [see panels \rm{II} and \rm{III} of Fig.~\ref{fig:mubmufDroplet}(b)]. 
Moreover for intermediate fermion densities, we see that the first-order boundary terminates in a manner reminiscent of a liquid-gas critical point. This remarkable quantum analog of liquid-gas critical behavior has also recently been predicted for BB mixtures~\cite{LiCriticality2023}.

\textit{Conclusions---}We have presented a novel variational ansatz for strongly interacting Bose-Fermi mixtures, and we have predicted the existence of quantum liquid-gas critical phenomena, including a unique self-bound droplet phase across the regime of unitarity-limited BF interactions. Our proposal is well within reach of current experiments. 
For instance, in a trapped system, the BF droplet could be observed by switching off the underlying trap and observing a stable self-bound state, similarly to droplets in Bose gases~\cite{Schmitt2016,FerrierBarbut2016, Chomaz2016,Cabrera2018,Semeghini2018}. %
On the other hand, in a uniform system such as an optical box potential, the system can be %
prepared with the appropriate density ratio but below the actual droplet density, after which it will spontaneously phase separate into regions of vacuum and regions of self-bound BF droplets. %

We emphasize that the existence of BF droplets relies on a combination of bosonic repulsion, Fermi pressure, and a sufficiently strong BF attraction. These conditions are not unique to ultracold atomic gases and could, for instance, be realized in charge-doped atomically thin semiconductor microcavities, where resonant attraction between electrons and exciton-polaritons (electron-hole bound state strongly coupled to light) was recently reported in the context of Fermi polarons~\cite{Sidler2017}. Thus, Bose-Fermi droplets are likely to exist also in platforms involving strong coupling of light and matter.

\acknowledgments 
We acknowledge useful discussions with Matteo Caldara, Francesca Maria Marchetti, Brendan Mulkerin, and Anthony Zulli. We also thank Anthony Zulli for help with figures, and Matteo Caldara and Pierbiagio Pieri for useful feedback on the manuscript. We acknowledge support from the Australian Research Council (ARC) Centre of Excellence in Future Low-Energy Electronics Technologies (CE170100039).  JL and MMP are also supported through the ARC Discovery Project DP240100569 and ARC Future Fellowship FT200100619, respectively, and jointly through the ARC Discovery Project
DP250103746. OB also acknowledges support from the Deutsche Forschungsgemeinschaft (DFG) via the Collaborative Research Centre SFB 1225 ISOQUANT (Project-ID No. 273811115). SF acknowledges support from an Australian Government Research Training Program (RTP) Scholarship.

\bibliography{ref}

\onecolumngrid
\newpage

\setcounter{equation}{0}
\setcounter{figure}{0}
\setcounter{table}{0}
\setcounter{page}{1}
\renewcommand{\theequation}{S\arabic{equation}}
\renewcommand{\thefigure}{S\arabic{figure}}

\begin{center}
\textbf{\large SUPPLEMENTAL MATERIAL:\\ ``Quantum droplets in a resonant Bose-Fermi mixture''}\\
\vspace{4mm}
S. Foster,$^1$
O. Bleu,$^{1,2}$
M. M. Parish$^1$,
and J. Levinsen$^1$\\
\emph{\small $^1$School of Physics and Astronomy, Monash University, Victoria 3800, Australia}\\
\emph{
$^2$Institut f{\"u}r Theoretische Physik, Universit{\"a}t Heidelberg, 69120 Heidelberg, Germany
}
\end{center}

\section{Model and ansatz}

For the calculations outlined here, we use the Hamiltonian from the main text, ${\mathcal{H}} = \mathcal{H}_0+ V_\mathrm{BF}+V_\mathrm{BB}$, with
\begin{subequations}
\begin{align}
    \mathcal{H}_0 &=  \sum_{\p} \bigg [
    (\epsilon_{{\p}}^\mathrm{F}-\mu_\mathrm{F}) f_{\p}^\dagger f_{\p}
    + (\epsilon_{\p}^\mathrm{B}-\mu_\mathrm{B}) b_{\p}^\dagger b_{\p}
    \bigg ], \\
    V_\mathrm{BF}&=g_\mathrm{BF} 
    \sum_{\p \p' \Q} f^\dagger_{\p} f_{\p'}
    b^\dagger_{\Q - \p}   b_{\Q-\p'},
    \\
    V_\mathrm{BB} &= \frac{g_\mathrm{BB} }{2}
    \sum_{\p \p' \Q} b^\dagger_{\p+\Q} 
    b^\dagger_{\p'-\Q}   b_{\p'}b_{\p}.
\end{align}
\end{subequations}
Here we have dispersions $\epsilon_{\p}^\mathrm{F/B} = |\p|^2/2m_\mathrm{{F/B}} \equiv p^2/2m_\mathrm{{F/B}}$ with masses $m_\mathrm{F/B}$, and 
we have set $\hbar$ and the system volume to 1. The weakly repulsive boson-boson interaction is treated at the mean-field level such that $g_\mathrm{BB}= 4\pi a_{\mathrm{BB}}/m_\mathrm{B}$, with scattering length $a_{\mathrm{BB}}>0$. The (bare) boson-fermion interaction strength $g_\mathrm{BF}$ is related to the boson-fermion scattering length $a_\mathrm{BF}$ via the renormalization condition
\begin{equation}\label{eq:BF-renormalization}
    \frac{1}{g_\mathrm{BF}} 
    =
    \frac{m_\mathrm{r}}{2\pi a_\mathrm{BF}} 
    -
    \sum_{\p}^{\Lambda}
    \frac{1}{\epsilon_{\p}^\mathrm{F} + \epsilon_{\p}^\mathrm{B}} \, ,
\end{equation}
where $m_\mathrm{r} = m_\mathrm{F} m_\mathrm{B}/(m_\mathrm{F} + m_\mathrm{B})$ and $\Lambda$ is an ultraviolet cutoff that will be sent to infinity at the end of the calculation. %
We additionally make use of the ansatz in Eq.~(2) of the main text, which we reproduce here for completeness
\begin{equation} \label{eq:ansatz-exp2}
    \ket{\psi} =\mathcal{N}e^{\alpha_0 b_0^\dag+\sum_{\k\q} \alpha_{\k\q} B_{\k \q}^\dag} \FS ,
\end{equation}

where the operator $B_{\k \q}^\dag\equiv b_{\q-\k}^\dag f_\k^\dag f_\q$, with $q\leq k_\mathrm{F}<k$ in Eq.~\eqref{eq:ansatz-exp2} and in the following.
These operators satisfy the commutation relations $[B^\dag_{\k\q},B^\dag_{\k'\q'}]=0$ and 
\begin{equation}  \label{eq:commutator2}
           [B_{\k' \q'}, B_{\k \q}^\dagger]
     \!=\! \delta_{\q \q'} \delta_{\k \k'} (1 - f_{\q} f^\dag_{\q} - f_\k^\dag f_\k + b^\dag_{\q-\k} b_{\q-\k}) - \delta_{\q \q'}  f^\dag_{\k} f_{\k'} b^\dag_{\q-\k} b_{\q-\k'} + f_\q f^\dag_{\q'} ( f^\dag_\k f_{\k'} \delta_{\q-\k,\q'-\k'} - b^\dag_{\q-\k} b_{\q'-\k} \delta_{\k\k'}).
\end{equation}
The following commutators are also useful:
\begin{subequations} \label{eq:commutators}
    \begin{align}
        &[f_{\p_1}^\dag f_{\p_2}, B^\dag_{\k\q}] 
        = b^\dag_{\q-\k} (\delta_{\k\p_2} f_{\p_1}^\dag f_\q
        - \delta_{\q\p_1} f^\dag_\k f_{\p_2}),
        \\
        &[b_{\p_1}^\dag b_{\p_2}, B^\dag_{\k\q}]
        =\delta_{\p_2, \q-\k} b_{\p_1}^\dag f^\dag_\k f_\q.
        \label{eq:bbCommutator}
    \end{align}
\end{subequations}

\subsection{Approximate normalization of the ansatz}

Since the operators in Eq.~\eqref{eq:ansatz-exp2} commute, we can rearrange terms such that we have a product over $\q$ as follows
\begin{equation}
    \ket{\psi}=
    \mathcal{N} e^{ \alpha_0 b_0^\dagger} \prod_{\substack{\q }}
    \bigg(
    1 + \sum_{\substack{\k}} \alpha_{\k \q} B_{\k \q}^\dag
    \bigg)\FS \equiv 
    \mathcal{N} e^{ \alpha_0 b_0^\dagger}
    \bigg(
    1 + \sum_{\substack{\k_1}} \alpha_{\k_1 \q_1} B_{\k_1 \q_1}^\dag
    \bigg) \bigg(
    1 + \sum_{\k_2} \alpha_{\k_2 \q_2} B_{\k_2 \q_2}^\dag
    \bigg)  \cdots \FS \, ,
\end{equation}
where each $\q$ is distinct, i.e., $\q_1 \neq \q_2 \neq \q_3 \ldots$. This form of the ansatz is convenient since the normalization then corresponds to
\begin{align} \nn
     1 & =  \braket{\psi} = \mathcal{N}^2 e^{|\alpha_0|^2} \\ \label{eq:norm}
     & \times
    \lFS \cdots \bigg(1 +\sum_{\k_2'} \alpha_{\k_2' \q_2}^* B_{\k_2' \q_2}
    \bigg)
    \underbrace{\bigg(1 +\sum_{\k_1'} \alpha_{\k_1' \q_1}^* B_{\k_1' \q_1}
    \bigg)
    \bigg(1 +  \sum_{\k_1} \alpha_{\k_1 \q_1} B_{\k_1 \q_1}^\dagger
    \bigg)}_{1+ \sum_{\k_1\k'_1}\alpha_{\k_1' \q_1}^*\alpha_{\k_1 \q_1}[B_{\k_1' \q_1}, B_{\k_1 \q_1}^\dagger]} \bigg(1 +  \sum_{\k_2} \alpha_{\k_2 \q_2} B_{\k_2 \q_2}^\dagger
    \bigg) \cdots \FS,
\end{align}
where we have rewritten the $\q_1$ terms using the fact that $B_{\k \q_1} B^\dag_{\k_2\q_2} B^\dag_{\k_3\q_3} \cdots \FS = 0$  due to Pauli exclusion since $\q_1 \neq \q_2, \q_3 \ldots$
To proceed, we must consider the commutator 
\begin{align} \label{eq:commutator3}
       [B_{\k' \q}, B_{\k \q}^\dagger]
    & \!=\! \delta_{\k \k'} (1 - f_\k^\dag f_\k + b^\dag_{\q-\k} b_{\q-\k}) - f^\dag_{\k} f_{\k'} b^\dag_{\q-\k} b_{\q-\k'} - f_\q f^\dag_{\q} (1- f^\dag_\k f_{\k} + b^\dag_{\q-\k} b_{\q-\k}) \delta_{\k\k'} ,
\end{align}  
where we have used Eq.~\eqref{eq:commutator2}.
First, we observe that the term involving $f_\q f^\dag_\q$ will give zero once inserted into Eq.~\eqref{eq:norm} since $[f_\q f^\dag_\q, B^\dag_{\k\q'}] = 0$ for $\q \neq \q'$, and $f_\q f^\dag_\q \FS = 0$.
We can then make use of the fact that the phase space available outside the Fermi sea is much larger than the phase space available inside (indeed, similar phase-space arguments have been used when calculating the interactions between Fermi polarons in the context of doped semiconductors~\cite{Tan2020}). Consequently, the expectation values of $-f_\k^\dag f_\k+ b^\dag_{\q-\k} b_{\q-\k}$ %
approximately vanish, since the number of bosons and fermions excited to momenta exceeding $k_\mathrm{F}$ are correlated and nearly equal. Finally, one can show that the remaining term $f^\dag_{\k}f_{\k'} b^\dag_{\q-\k} b_{\q-\k'}$ is only nonzero if there are two or more $B^\dag$ operators to the right of it (or two or more $B$ operators to the left of it) in Eq.~\eqref{eq:norm}, which is once again due to each $\q$ being distinct. This implies a correlation involving three fermions within the restricted phase space $\q_1-\q_2 = \k_2 - \k_3 \neq 0$, which is expected to be small due to approximate cancellations arising from Fermi statistics.

Therefore, we can effectively take $[B_{\k' \q}, B_{\k \q}^\dagger] \simeq \delta_{\k\k'}$ in Eq.~\eqref{eq:norm}, such that we obtain
\begin{equation}
    1 =  \braket{\psi}  \simeq \mathcal{N}^2 e^{|\alpha_0|^2} \prod_\q \bigg(1+\sum_\k |\alpha_{\k\q}|^2 \bigg) \braket{\rm FS} ,
\end{equation}
which corresponds to the approximate normalization constant used in the main text: %
\begin{equation}
    \mathcal{N} \simeq
    \frac{e^{- |\alpha_0|^2/2}}{\prod_{\q} \sqrt{1 + \sum_{\k} |\alpha_{\k \q}|^2}}.
\end{equation}
This approximation is used to compute all the expectation values.

\section*{Free Energy}

The free energy corresponds to $\Omega = \bra{\psi}\mathcal{H}\ket{\psi} \equiv \expval{\mathcal{H}}$, the evaluation of which requires the expectation values of the fermion and boson number operators, as well as those of the BF and BB interactions. To evaluate these, we relate each expectation value to an expectation value of the operator $B_{\k\q}^\dag$. We can then easily compute the expectation values by defining the ket $\ket{\phi}\equiv e^{\sum_{\k\q} \alpha_{\k\q} B_{\k\q}^\dag}\FS$ corresponding to the fluctuation part of the state $\ket{\psi}$, and then recognizing that 
\begin{align}    
\expval*{B^\dagger_{\k \q}} = \mathcal{N}^2 e^{|\alpha_0|^2}\bra{\phi} B^\dag_{\k\q}\ket{\phi} =  \mathcal{N}^2 e^{|\alpha_0|^2}\frac{\partial \expval*{\phi|\phi}}{\partial \alpha_{\k\q}}.
\end{align}
This gives
\begin{equation}\label{eq:Bdagexpval}
    \expval*{B^\dagger_{\k\q}}=
    \frac{\alpha_{\k \q}^*}
    {1+\sum_{\k'}|\alpha_{\k' \q}|^2},
\end{equation}
where we have used the same approximate normalization as above for $\ket{\phi}$, i.e., $\expval{\phi|\phi} = \prod_{\q} (1 + \sum_\k |\alpha_{\k\q}|^2)$.
In the following, we will also repeatedly make use of the commutation relation:
\begin{equation} \label{eq:exp_commutator}
    [ \hat{X},e^{ \hat{Y}} ] = e^{ \hat{Y}} [\hat{X},\hat{Y}]   + \frac{1}{2} e^{ \hat{Y}} [[\hat{X},\hat{Y}],\hat{Y}] ,
\end{equation}
which holds for operators $\hat X$ and $\hat Y$ that satisfy $[[[\hat X,\hat Y],\hat{Y}],\hat{Y}]=0$.

Throughout the remainder of the supplementary material, the variational parameters are assumed to be real without loss of generality.
 
\subsection*{Boson and fermion occupations}

To evaluate the single-particle terms, we first obtain the condensate density:
\begin{align}
    n_0 = \expval*{b_0^\dag b_0}=\alpha_0^2.
\end{align}
We then determine the other occupations by noting that the commutators in Eq.~\eqref{eq:commutators} both 
commute with $B_{\k' \q'}^\dag$, such that we can
use Eq.~\eqref{eq:exp_commutator}. For example, the number of fermions with momentum $p>k_\mathrm{F}$ is
\begin{equation}
    \expval*{f^\dagger_{\p} f_{\p}}_{p>k_\mathrm{F}} = \mathcal{N}^2 e^{\alpha_0^2}
    \lFS e^{ \sum_{\k' \q'} \alpha_{\k' \q'} B_{\k' \q'} } 
    f^\dagger_{\p} f_{\p}
    e^{\sum_{\k \q} \alpha_{\k \q} B^\dagger_{\k \q} } \FS.
\end{equation}
Then Eq.~\eqref{eq:exp_commutator} yields 
\begin{equation}
    [f^\dagger_{\p} f_{\p}, e^{\sum_{\k \q} \alpha_{\k \q} B^\dagger_{\k \q} } ] =  
    \sum_{\q'} \alpha_{\p \q'} B^\dag_{\p\q'}
    e^{\sum_{\k \q} \alpha_{\k \q} B^\dagger_{\k \q} },
\end{equation}
which is used to rewrite the number operator expectation value in terms of the $B_{\k\q}^\dag$ operator 
\begin{equation}
    \expval*{f^\dagger_{\p} f_{\p}}_{p>k_\mathrm{F}} = \mathcal{N}^2 e^{\alpha_0^2}
    \sum_{\q''} \alpha_{\p\q''}
    \lFS e^{ \sum_{\k' \q'} \alpha_{\k' \q'} B_{\k' \q'} } 
    B_{\p\q''}^\dag
    e^{\sum_{\k \q} \alpha_{\k \q} B^\dagger_{\k \q} } \FS.
\end{equation}
A similar process can be applied to the other occupations, which finally results in
\begin{subequations}
\begin{align}
    \expval*{f_{\p}^\dagger f_{\p} }_{p>k_\mathrm{F}} &= 
    \sum_{\q}  \alpha_{\p \q} \expval*{B_{\p \q}^\dag}
    = \sum_\q \frac{\alpha_{\p\q}^2}{1+\sum_{\k}\alpha_{\k\q}^2}
    ,
    \\
    \expval*{f_{\p}^\dagger f_{\p} }_{p<k_\mathrm{F}}
     &=
     1 - \sum_{\k} \alpha_{\k \p} \expval*{B_{\k \p}^\dag}
     =1-\sum_\k \frac{\alpha_{\k\p}^2}{1+\sum_{\k'}\alpha_{\k'\p}^2},
     \\
     \expval*{b_{\p}^\dagger b_{\p} }_{p\neq0}&=
     \sum_{\k \q} \delta_{\p, \q -\k} \alpha_{\k \q}\expval*{B^\dagger_{\k \q}}
     =
     \sum_{\k\q} \delta_{\p,\q-\k} \frac{\alpha_{\k\q}^2}{1+\sum_{\k'}\alpha_{\k'\q}^2}
     .
\end{align}
\end{subequations}
where in the last step of each line we have used  Eq.~\eqref{eq:Bdagexpval}. In particular, we see that $\sum_\p \expval{f_{\p}^\dagger f_{\p}} = \sum_{|\p| <k_\mathrm{F}} = n_\mathrm{F}$, as required.

Therefore, the expectation value of the non-interacting part of the Hamiltonian is
\begin{equation}\label{eq:H0SM}
     \expval*{\mathcal{H}_0}=
    \sum_{\k \q} (\epsilon_{\k}^\mathrm{F}-\epsilon_{\q}^\mathrm{F}+\epsilon_{\q-\k}^\mathrm{B}-\mu_\mathrm{B}) v_{\k \q}^2
    +
        \sum_{\q} ( \epsilon_{\q}^\mathrm{F} - \mu_\mathrm{F} )
    -\mu_\mathrm{B} \alpha_0^2.
\end{equation}
Here, as in the main text, we have %
\begin{align}
u_{\q} 
    &=  \frac{1}{\sqrt{1+\sum_{\k'}|\alpha_{\k'\q}|^2}}, %
\end{align}
and
\begin{align}
v_{\k\q}&=  \alpha_{\k\q}u_\q.
\end{align}

\subsection{Boson-fermion interaction}

With respect to the Bose-Fermi interaction term, we first separate out the terms involving the condensate:
\begin{align} \label{eq:VBF0}
    \expval*{V_\mathrm{BF}}
    &= g_\mathrm{BF}\sum_{\substack{\p , \p' , \Q \\ \p \neq \Q \\ 
    \p' \neq \Q 
    }} \expval*{f^\dagger_{\p} f_{\p'}
    b^\dagger_{\Q - \p}   b_{\Q-\p'}}
    +
    g_\mathrm{BF} \alpha_0\sum_{\k \q}\expval*{B_{\k\q}^\dagger} 
    +g_\mathrm{BF} \alpha_0\sum_{\k \q}\expval*{B_{\k\q}} 
    +g_\mathrm{BF} \alpha_0^2\sum_{\p} \expval*{f_{\p}^\dagger f_{\p}}.
\end{align}
Here we have used the fact that $\expval*{b^\dagger_{\p' - \p} f^\dagger_{\p} f_{\p'}}$ is only nonzero if $p'\leq k_\textrm{F} < p$ since there are otherwise more boson excitations than particle-hole excitations, which gives zero in our ansatz. 
We have already computed the expectation values in the last three terms, so we are only left with evaluating the first term in Eq.~\eqref{eq:VBF0}.  

Since this first term is independent of the BEC, we only need to consider how the operators act on the state $\ket{\phi}$. Note that the expectation value is only nonzero if $p,p' < \kf$ or $p,p' > \kf$, since we otherwise create more particle-hole excitations than boson excitations. 
Using Eqs.~\eqref{eq:commutators} and \eqref{eq:exp_commutator} then
gives
\begin{align}
\sum_{\substack{\p , \p' , \Q \\ \p \neq \Q \\ 
    \p' \neq \Q 
    }} \expval*{f^\dagger_{\p} f_{\p'}
    b^\dagger_{\Q - \p}   b_{\Q-\p'}}
    &=
     \sum_{\substack{\k_1 \q_1 \\ \k_2 \k' \q_2}} \alpha_{\k_1 \q_1} \alpha_{\k_2 \q_2}
    \expval*{b^\dag_{\q_1 - \k_1 -\k' + \k_2} f^\dag_{\k_1} f_{\q_1} b^\dag_{\q_2-\k_2} f^\dag_{\k'}f_{\q_2}}
    \nn \\&
    -\sum_{\substack{\k_1 \q_1 \\ \k_2 \q_2 \q'}} \alpha_{\k_1 \q_1} \alpha_{\k_2 \q_2}
    \expval*{b^\dag_{\q_1 - \k_1 -\q_2+\q'} f^\dag_{\k_1} f_{\q_1} b^\dag_{\q_2-\k_2} f^\dag_{\k_2 }f_{\q' }}
    \nn \\
    &+ n_\mathrm{F} \sum_{\k\q} \alpha_{\k\q} \expval*{B^\dag_{\k \q}} 
     +\sum_{\k \k' \q} \alpha_{\k \q}
    \expval*{B^\dag_{\k' \q}}
    -\sum_{\k \q \q'} \alpha_{\k \q}
    \expval*{B^\dag_{\k \q'}} \, .
\end{align}

The terms in the last line can be straightforwardly computed using Eq.~\eqref{eq:Bdagexpval}, while the first two terms correspond to higher-order correlations involving multiple fermions. We can show that these correlations approximately cancel due to the Fermi statistics of boson-fermion pairs. For instance, if we expand the wave function in the first two terms such that we consider two $B$ operators, then we have
\begin{align} \nn
     & \sum_{\substack{\k_1 \k_1' \q_1 \\ \k_2 \k_2' \q_2}} \alpha_{\k_1 \q_1} \alpha_{\k_2\q_2} \alpha_{\k_1' \q_1} \alpha_{\k_2' \q_2} 
   \left( \bra{\mathrm{FS}} B_{\k_2'\q_2} B_{\k_1'\q_1}\sum_{\k'}b^\dag_{\q_1 - \k_1 -\k' + \k_2} f^\dag_{\k_1} f_{\q_1} b^\dag_{\q_2-\k_2} f^\dag_{\k'}f_{\q_2}\FS \right.  \\ \nn
   & \hspace{50mm} \left. - \bra{\mathrm{FS}} B_{\k_2'\q_2} B_{\k_1'\q_1} \sum_{\q'} b^\dag_{\q_1 - \k_1 -\q_2+\q'} f^\dag_{\k_1} f_{\q_1} b^\dag_{\q_2-\k_2} f^\dag_{\k_2 }f_{\q' } \FS \right) \\ \nn
   & =  \sum_{\substack{\k_1 \k_2 \\ \q_1 \q_2}} \alpha_{\k_1\q_1} \alpha_{\k_2\q_2}\left[\delta_{\q_1-\q_2, \k_1-\k_2} \alpha_{\k_1\q_1}\sum_{\k'} \alpha_{\k'\q_2} -  \delta_{\k_1\k_2} \alpha_{\k_1\q_2}  \sum_{\k'} \alpha_{\k'\q_1}  \right. \\
   & \hspace{10mm} \left. - \, \alpha_{\k_1\q_2} \sum_{\k'} \alpha_{\k'\q_1} \delta_{\q_1-\q_2, \k'-\k_2} + \alpha_{\k_2\q_1}  \sum_{\q'} \alpha_{\k_1\q'} \delta_{\q'-\q_2, \k_1-\k_2} - \delta_{\q_1-\q_2, \k_1-\k_2} \alpha_{\k_1 \q_1} \sum_{\q'} \alpha_{\k_2\q'} \right] \, .
\end{align}
The last three terms are found to vanish when we take 
the cutoff $\Lambda\to\infty$ and  $g_\mathrm{BF}\to0$, since
$\alpha_{\k\q}$ scales as $1/k^2$ for large $k$ 
and the sums are convergent. On the other hand, the first two terms involving sums over $\k'$ diverge as $\Lambda\to\infty$ and thus yield a finite contribution when multiplied by $g_\mathrm{BF}$. However, the two terms approximately cancel each other for $k \gg q$ and hence we can neglect them. 

Thus, the total Bose-Fermi interaction is
\begin{align}
    \expval*{V_\mathrm{BF}}
    \simeq & \,
    g_\mathrm{BF} \sum_{\k \k' \q} \alpha_{\k \q} \expval*{B^\dag_{\k' \q}}
    -
    g_\mathrm{BF} \sum_{\k \q \q'} \alpha_{\k \q} \expval*{B^\dag_{\k \q'}} + n_\mathrm{F} g_\mathrm{BF} \sum_{\k\q} \alpha_{\k\q} \expval*{B^\dag_{\k\q}}
    \nn \\
    &+
    g_\mathrm{BF}\alpha_0 \sum_{\k \q}\expval*{B_{\k\q}^\dagger} 
    +g_\mathrm{BF}\alpha_0 \sum_{\k \q}\expval*{B_{\k\q}} 
    +
    g_\mathrm{BF} \alpha_0^2 \sum_{\p} \expval*{f^\dag_\p f_\p}
\end{align}
Upon substituting in the known expectation values we find that
\begin{align}
    \expval*{V_\mathrm{BF}}
    & = 
    2g_\mathrm{BF}\alpha_0 \sum_{\k \q} u_{\q} v_{\k \q}
    + 
    g_\mathrm{BF}
    \sum_{\k\q} v_{\k\q} \left( \sum_{\k'} v_{\k'\q} - \sum_{\q'}v_{\k\q'} \right)
    +
     g_\mathrm{BF} n_\mathrm{F} \Big(\alpha_0^2 +\sum_{\k\q} v_{\k\q}^2 \Big) \, .
\end{align}
Here, the last term vanishes when we take the cutoff $\Lambda\to\infty$ and  $g_\mathrm{BF}\to0$ since $\alpha_0^2 +\sum_{\k\q} v_{\k\q}^2$ is finite. Likewise, we can drop the sum over $\q'$ which is subleading relative to the sum over $\k'$. As noted above, this latter sum formally diverges as $\Lambda\to\infty$ since %
$\vkq$ scales as $1/k^2$ for large $k$ (see, also, the equations of motion in the next section). 

This finally yields the following expression for the boson-fermion interaction 
\begin{equation}\label{eq:VBFSM}
    \expval*{V_{\mathrm{BF}}}= 2g_\mathrm{BF}\alpha_0 \sum_{\k \q} u_{\q} v_{\k \q}
    +g_\mathrm{BF} \sum_{\k \k' \q} v_{\k \q} v_{\k' \q} - g_\mathrm{BF} \sum_{\k \q \q'} v_{\k \q} v_{\k \q'}.
\end{equation}

\subsection{Boson-boson interaction}

The Bose-Bose interaction can also be simplified within the ansatz. Once again, we decompose the interaction in terms of condensed and uncondensed bosons:
\begin{equation}
    \expval*{V_\mathrm{BB}} = \frac{g_\mathrm{BB}}{2} \alpha_0^4%
    +
    2 g_\mathrm{BB} \alpha_0^2 %
    \sum_{\p\neq 0} \expval*{b_{\p}^\dag b_{\p}}
    +
\frac{g_\mathrm{BB} }{2}
\sum_{\substack{\p,\p',\Q \\ 
   \p, \p' \neq \0 \\ 
    \Q \neq -\p, \p' 
    }}
    \expval*{b^\dagger_{\p+\Q} 
    b^\dagger_{\p'-\Q}   b_{\p'}b_{\p}}
\end{equation}
The correlator in the last term approximately vanishes due to Fermi statistics once again, since we can rewrite it as
\begin{align} \nn
    \sum_{\substack{\p,\p',\Q \\ 
   \p, \p' \neq \0 \\ 
    \Q \neq -\p, \p' 
    }}
    \expval*{b^\dagger_{\p+\Q} 
    b^\dagger_{\p'-\Q}   b_{\p'}b_{\p}} & =  \sum_{\substack{\k_1 \k_2 \Q \\ \q_1 \q_2}} \alpha_{\k_1 \q_1} \alpha_{\k_2 \q_2}
    \expval*{b^\dag_{\q_1 - \k_1 +\Q} f^\dag_{\k_1} f_{\q_1} b^\dag_{\q_2-\k_2-\Q} f^\dag_{\k_2}f_{\q_2}} \\
   &  =  \frac{1}{2} \sum_{\substack{\k_1 \k_2 \Q \\ \q_1 \q_2}} (\alpha_{\k_1 \q_1} \alpha_{\k_2 \q_2} - \alpha_{\k_2 \q_1} \alpha_{\k_1 \q_2}) 
    \expval*{b^\dag_{\q_1 - \k_1 +\Q} f^\dag_{\k_1} f_{\q_1} b^\dag_{\q_2-\k_2-\Q} f^\dag_{\k_2}f_{\q_2}} \, ,
\end{align}
where we have used Eqs.~\eqref{eq:bbCommutator} and \eqref{eq:exp_commutator} in the first line, and we have exchanged fermion operators in the second line, relabelling and redefining dummy variables.

Thus, using the expectation values that have already been computed above, we finally obtain
\begin{equation}\label{eq:VBBSM}
    \expval*{V_\mathrm{BB}}= \frac{g_\mathrm{BB}}{2} \alpha_0^4
    + 2 g_\mathrm{BB} \alpha_0^2
    \sum_{\k \q} v_{\k \q}^2 .
\end{equation}

\subsection{Total free energy} %

The free energy, $\Omega = \expval{\mathcal{H}_0}+ \expval{V_\mathrm{BF}}+\expval{V_\mathrm{BB}}$, can now be obtained by collecting all the terms from Eqs.~\eqref{eq:H0SM}, \eqref{eq:VBFSM}, and \eqref{eq:VBBSM} above, giving
\begin{align}
    \Omega
    = 
    \sum_{\q} ( \epsilon_{\q}^\mathrm{F} - \mu_\mathrm{F} )
    -\mu_\mathrm{B} n_0
    +
    \frac{g_\mathrm{BB}}{2} n_0^2
    +\sum_{\k \q} E_{\k \q}v_{\k \q}^2
    +
    2g_\mathrm{BF}\sqrt{n_0}\sum_{\k \q} u_{\q} v_{\k \q}     +g_\mathrm{BF} \sum_{\k \k' \q} v_{\k \q} v_{\k' \q}
    \label{eq:freeenergySM}
\end{align}
As in the main text we define
\begin{equation}
    {E}_{\k\q} \equiv\epsilon_{\k}^\mathrm{F}-\epsilon_{\q}^\mathrm{F}+\epsilon_{\q-\k}^\mathrm{B}-\mu_\mathrm{B} + 2 g_\mathrm{BB}n_0,
    \label{eq:SM-Ekq}
\end{equation}
with boson density $n_\mathrm{B} = n_0 + \sum_{\k \q} v_{\k\q}^2$. 

\section{Equations of Motion and variational parameters}
The free energy in Eq.~\eqref{eq:freeenergySM} leads to the equation of motion for the fluctuations
\begin{align}
    &    E_{\k \q} v_{\k \q}=
    (\sqrt{n_0} v_{\k \q}-u_{\q})
    g_{\mathrm{BF}} \sum_{\k'} \frac{v_{\k' \q}}{u_\q}
    -
    g_{\mathrm{BF}}\sqrt{n_0} u_{\q},
    \label{eq:SM-min-akq}
\end{align}
where we have used $\partial u_\q / \partial v_{\k\q} = - v_{\k\q}/u_\q$ which follows from the condition $u_\q^2 + \sum_{\q} \vkq^2 = 1$. Equation~\eqref{eq:SM-min-akq} is then used to evaluate the free energy in Eq.~\eqref{eq:freeenergySM} at the minimum with respect to fluctuations, which yields
\begin{align}
       \Omega =&
    \frac{3}{5} E_\mathrm{F} n_\mathrm{F}-\mu_\mathrm{F}n_\mathrm{F}
    -\mu_\mathrm{B}n_0 
    + \frac{g_{\mathrm{BB}} }{2} n_0^2 +n_0 \sum_{\q} \chi_{\q} , 
    \label{eq:SM-free-energy-renormalised}
\end{align}
where we have defined
\begin{align}\label{eq:chiSM}
    \chi_{\q} \equiv
    \frac{g_{\mathrm{BF}}}{\sqrt{n_0}} \sum_{\k} \frac{v_{\k\q}}{u_\q}
    =
    \Bigg[\frac{1}{g_{\mathrm{BF}}}
    +
    \sum_{\substack{\k}}
    \frac{1}{E_{\k\q}   - n_0 \chi_\q}
    \Bigg]^{-1}.
\end{align}
The last step follows by noting that Eq.~\eqref{eq:SM-min-akq} can be rearranged for the ratio between the variational parameters
\begin{equation}\label{eq:vuratioSM}
    \frac{v_{\k \q}}{u_{\q}}=
    \frac{\sqrt{n_0}(\chi_{\q} +g_\mathrm{BF})}{n_0 \chi_{\q} -E_{\k \q}}
    ,
\end{equation}
combined with the result $g_\mathrm{BF}\sum_\k\frac1{n_0\chi_\q-E_{\k\q}}\to1$ within the renormalization procedure. We may also obtain the functions $u_\q$ and $v_{\k\q}$ in terms of $\chi_\q$ by imposing the normalization condition $u_{\q}^2 + \sum_{\k} v_{\k \q}^2=1$ and using Eq.~\eqref{eq:vuratioSM} together with the renormalization condition for $g_\mathrm{BF}$. This gives
\begin{equation}
    u_{\q}^2
    =
    \frac{1}
    {1 + n_0 \chi_{\q}^2 
    \sum_{\k} (n_0 \chi_{\q} -  E_{\k \q})^{-2}},
\end{equation}
and 
\begin{equation}\label{eq:SM-vkq2}
    v_{\k \q}^2 =
    \frac{n_0 \chi_{\q}^2}{(n_0 \chi_{\q} - E_{\k \q})^2}
    u_{\q}^2 .
\end{equation}

We then obtain the additional equations of motion from the free energy in Eq.~\eqref{eq:SM-free-energy-renormalised}
\begin{subequations}
    \begin{align}         &\mu_\mathrm{B} = 
    g_{\mathrm{BB}} (2n_\mathrm{B} -n_0)+
    \sum_{\q} u_{\q}^2 \chi_\q 
   ,
    \label{eq:SM-min-n0}
    \\
        &\mu_\mathrm{F} = 
    E_\mathrm{F} + n_0 \chi_{k_\mathrm{F}}
    + \sum_{\q} (E_{k_\mathrm{F} \q} - n_0 \chi_{\q})
    v_{k_\mathrm{F} \q }^2 ,
    \label{eq:SM-min-kf}
    \end{align}
\end{subequations}
where we have used the expressions for $u_\q$, $v_{\k\q}$ and $\chi_\q$. In practice, we calculate all the parameters by solving Eq.~\eqref{eq:chiSM} for $\chi_\q$ self-consistently.

\begin{figure}[th]
    \centering
    \includegraphics[width=0.5\linewidth]{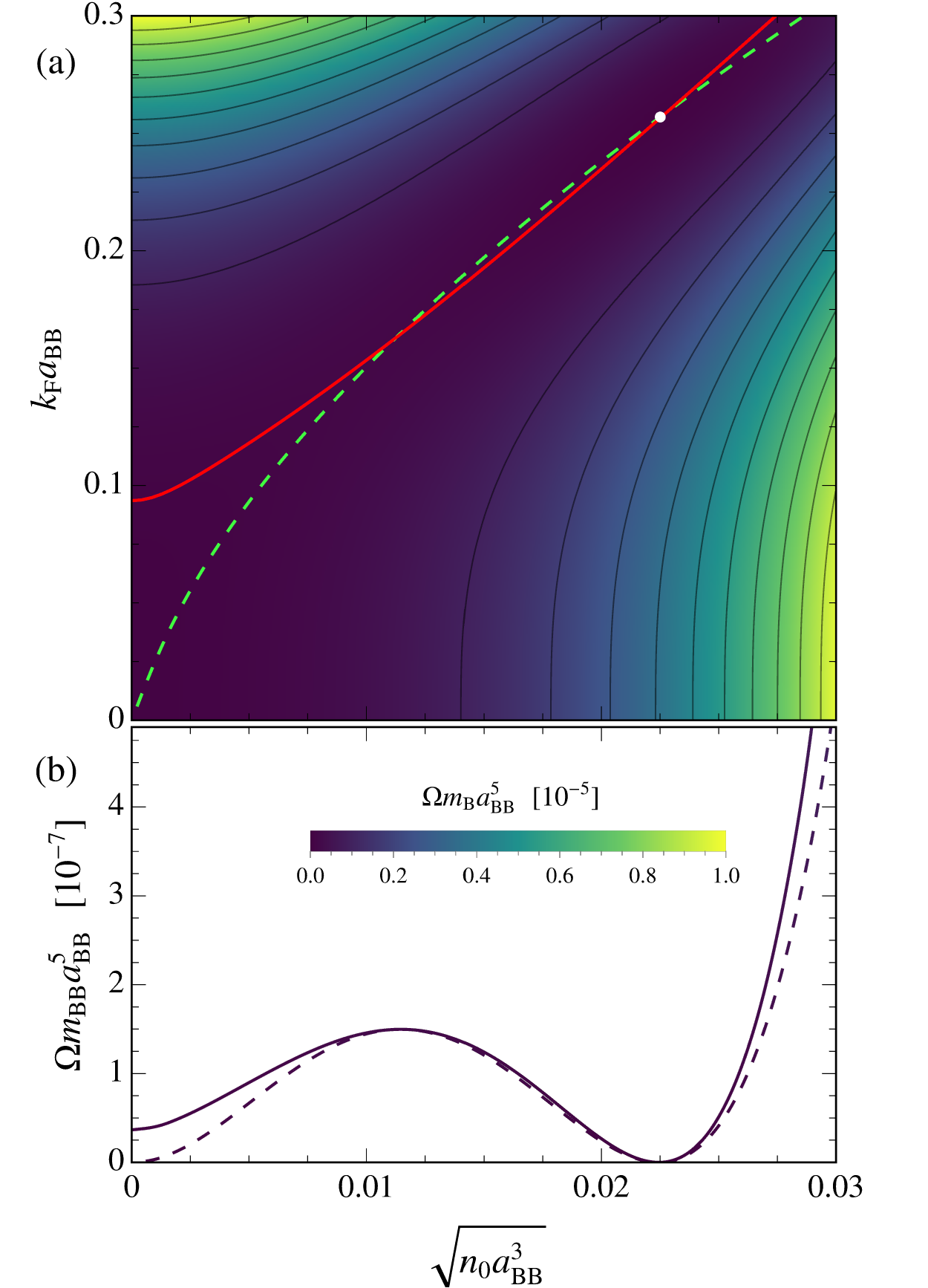}
    \caption{(a) Free energy as a function of the condensate density and the Fermi momentum for parameters $a_\mathrm{BB}/a_\mathrm{BF} = 0$, $\mu_\mathrm{F}=0$, $m_\mathrm{B}/m_\mathrm{F}=23/40$ and $\mu_\mathrm{B}/\epsilon_\mathrm{BB} = -2.077\times 10^{-3}$. Overlayed are the equations of motion  $\partial\Omega/\partial n_0=0$ and $\partial\Omega/\partial n_\mathrm{F}=0$  giving the  red, solid  and green dashed line, respectively. The white dot indicates the droplet phase. (b) The free energy curve as a function of the condensate density plotted along the minimization conditions. The solid~(dashed) line corresponds to the solid~(dashed) equation of motion in panel (a). 
    }
    \label{fig:dropletContour}
\end{figure}

\section*{Free-Energy Minimum}
In order to ensure that the stationary conditions given by our equations of motion are minima, we must have a positive curvature. We first examine whether $\partial^2\Omega/\partial v_{\k\q}^2>0$ is satisfied, since we use it in simplifying the free energy expression before determining the remaining equations of motion. Starting from the free energy in Eq.~\eqref{eq:freeenergySM}, i.e., before applying the minimization with respect to fluctuations, we obtain
\begin{equation}
    \frac{\partial^2 \Omega}{\partial v_{\k\q}^2} = 2(E_{\k\q} - n_0 \chi_\q)
    - 2n_0 \chi_\q \frac{v_{\k\q}^2}{u_\q^2}.
\end{equation}
Since we have $E_{\k\q} > 0$ (which follows from $k>q$ and $\mu_\mathrm{B} < 0$), and since we find that $\chi_\q<0$ within the regime investigated, we always have $\partial^2\Omega/\partial v_{\k\q}^2>0$, thus ensuring that we are at a minimum with respect to fluctuations. %

We then use the free energy evaluated at the minimum of $v_{\k\q}$, Eq.~\eqref{eq:SM-free-energy-renormalised}, to obtain $\Omega$ as a function of the remaining parameters, $n_\mathrm{F}$ and $n_0$. As this $\Omega$ is already a minimum in $\vkq$, the ground state is obtained by simultaneously satisfying the two associated equations of motion, Eqs.~\eqref{eq:SM-min-n0} and~\eqref{eq:SM-min-kf}, and ensuring that we have minima and not maxima. Figure~\ref{fig:dropletContour} shows a typical free energy contour for the droplet phase, with parameters relevant to a Na-K mixture at unitarity, where we have overlaid the equations of motion. We see that the global minimum of $\Omega$ at finite densities---white dot of Fig.~\ref{fig:dropletContour}(a)---corresponds to the intersection of the equations of motion, and that it is indeed a droplet phase since it has $\Omega=0$.
The free energy along the  $n_0$ and $n_\mathrm{F}$ equations of motion are plotted as a function of the condensate density in Fig.~\ref{fig:dropletContour}(b) highlighting again how the droplet phase coincides with the intersection of the two equations of motion at $\Omega=0$.

\section*{Polaron Limits}

\subsection{Fermi polaron}

The Fermi polaron plays an important role in our ansatz, representing the large imbalance limit ($n_\textrm{B}/n_\textrm{F} \ll 1$) of a BF mixture. Specifically, the boson chemical potential, $\mu_\mathrm{B}$, associated with the single boson impurity limit corresponds to a spinodal line present in Fig.~\ref{fig:mubmufDroplet}. Here we briefly outline the calculation of the boson chemical potential  in this limit.

We begin with the general equations of motion, Eqs.~\eqref{eq:SM-min-akq} and \eqref{eq:SM-min-n0}. We then take the limit of $n_\mathrm{B}\to0$ which is equivalent to only considering the leading order in the original variational parameter, $\alpha_{\k\q}$. This can be seen from Eq.~\eqref{eq:ansatz-exp} which recovers the Chevy ansatz in this limit~\cite{Chevy2006}. Therefore, we perform the transformation $u_\q\simeq \sqrt{1 - \sum_{\k}\alpha_{\k\q}^2}$ and $v_{\k\q}\simeq \alpha_{\k\q}$ resulting in the Fermi polaron equations of motion
\begin{subequations}
    \begin{align}
    \mu_\mathrm{B} \alpha_0 &= g_{\mathrm{BF}}\sum_{\k\q} \alpha_{\k\q},
    \\
    \mu_\mathrm{B} \alpha_{\k\q}
    &=
    (\epsilon_\k^\mathrm{F} -\epsilon_\q^\mathrm{F} +\epsilon_{\q-\k}^\mathrm{B}  ) \alpha_{\k\q}
    + g_\mathrm{BF} \sum_{\k'}\alpha_{\k'\q} + g_\mathrm{BF} \alpha_0,
    \end{align}
\end{subequations}
which are equivalent to the equations of motion if we had used the Chevy ansatz for a single bosonic impurity immersed in a Fermi sea~\cite{Chevy2006}.

As the Fermi polaron energy plays the role of a spinodal line in Fig.~\ref{fig:mubmufDroplet}, we demonstrate how to obtain this curve from the equations of motion.
We first define $\chi_\q^{(0)}$ in the polaron limit:
\begin{equation}
    \chi^{(0)}_\q = \frac{g_\mathrm{BF}}{\alpha_0} \sum_{\k}
    \alpha_{\k\q} = 
    \Bigg[
    \frac{1}{g_\mathrm{BF}}
    +
    \sum_{\k} \frac{1}{E_{\k\q}}
    \Bigg]^{-1},
\end{equation}
where $E_{\k\q}$ is given by Eq.~\eqref{eq:SM-Ekq} but with $n_\mathrm{B} \to 0$. Then the equation of motion for the boson chemical potential reduces to
\begin{equation}
    \mu_\mathrm{B} = \sum_{\q}\chi^{(0)}_\q. 
\end{equation}
This equation, which is equivalent to using a non-self-consistent $T$-matrix~\cite{Prokofev2008}, must be solved for $\mu_\mathrm{B}$ to obtain the polaron energy.

In order to produce the dot-dashed/solid black line displayed in Fig.~\ref{fig:mubmufDroplet}, we relate the Fermi momentum to the fermion chemical potential using $\mu_\mathrm{F}=k_\mathrm{F}^2/2m_\mathrm{F}$. This follows from the limit $n_\mathrm{B}, n_0\to0$ where the fermion chemical potential, Eq.~\eqref{eq:SM-min-kf} reduces to the Fermi energy. Here we see from the ansatz, Eq.~\eqref{eq:ansatz-bcs-form}, that without any bosons to interact with we have $\vkq\to0$, thus leaving the bare Fermi sea.

\subsection{Bose polaron}
We may likewise recover the equations of motion for the Bose polaron---i.e., a single fermion impurity immersed in a BEC---by considering the limit of $n_\mathrm{F}\to0$, where the fermion chemical potential plays the role of the polaron energy. We start with equations~\eqref{eq:SM-min-kf} and \eqref{eq:SM-min-akq} and consider the constant order of the Taylor expansion in $n_\mathrm{F}$ to find that 
\begin{subequations}
\begin{align}
    &\mu_\mathrm{F} u_0 = g_\mathrm{BF} \sqrt{n_0} 
    \sum_{\p} v_\p,\label{eq:bosepolaron1}
    \\
    & 
    E_\p v_\p u_0 = \sqrt{n_0} v_\p g_\mathrm{BF}
    \sum_{\p'} v_{\p'}
    - g_\mathrm{BF} \sum_{\p'} v_{\p'} u_0
    - g_\mathrm{BF} \sqrt{n_0} u_0^2,
\end{align}
\end{subequations}
where we define $E_\p \equiv E_{\p0}$ and $v_{\p}\equiv v_{\p0}$.
It is common to see the second of these equations written as 
\begin{equation}\label{eq:bosepolaron2}
    \mu_\mathrm{F} v_\p = E_\p v_\p 
    + g_\mathrm{BF} \sum_{\p'} v_{\p'}
    + g_\mathrm{BF} \sqrt{n_0} u_0,
\end{equation}
which is obtained by substituting the first expression into the second one.

The equations of motion in Eqs.~\eqref{eq:bosepolaron1} and \eqref{eq:bosepolaron2} are then equivalent to the minimization conditions following the use of a variational ansatz that allows a single Bogoliubov excitation of the BEC, see Ref.~\cite{Li2014BosePolaron}.

\section*{Weak-coupling limit}

Here we demonstrate that our model reproduces the ground state energy to second order in the Bose-Fermi coupling constant, first obtained in Ref.~\cite{Viverit2002}. The starting point is the free energy of our model, Eq.~\eqref{eq:SM-free-energy-renormalised}. 

We relate this to the energy density $\mathcal{E} \equiv \Omega + \mu_\mathrm{B} n_\mathrm{B} + \mu_\mathrm{F} n_\mathrm{F}$ resulting in 
\begin{align}
    \mathcal{E} = \frac{3}{5} E_\mathrm{F} n_\mathrm{F} + 
    \mu_\mathrm{B} (n_\mathrm{B}- n_0) + n_0 \sum_{\q} \chi_\q
    + \frac{g_\mathrm{BB}}{2} n_0^2 
\end{align}
Defining $\delta n\equiv n_\mathrm{B}-n_0 = \sum_{\k\q} v_{\k\q}^2$, and using the mean-field chemical potential $\mu_\mathrm{B} = g_\mathrm{BB} n_\mathrm{B} + (2\pi a_\mathrm{BF}/m_r) n_\mathrm{F}$, we can rewrite the energy density as 
\begin{align} \label{eq:EdensSM}
    \mathcal{E} = \frac{3}{5} E_\mathrm{F} n_\mathrm{F} + 
    \bigg( g_\mathrm{BB} n_\mathrm{B} + \frac{2\pi a_\mathrm{BF}}{m_r} n_\mathrm{F}
    \bigg)
    \delta n + (n_\mathrm{B}-\delta n) \sum_{\q} \chi_\q
    + \frac{g_\mathrm{BB}}{2} [n_\mathrm{B}^2 - 2 n_\mathrm{B} \delta n + \delta n ^2].
\end{align}

We may now expand in the BF coupling constant. First, 
we note that $v_{\k\q}^2 \simeq n_0 \chi_\q^2/E_{\k\q}^2$ from Eq.~\eqref{eq:SM-vkq2}, which thus yields the depletion
\begin{equation}
    \delta n \simeq n_\mathrm{B} \bigg(\frac{2\pi a_\mathrm{BF}}{m_r}\bigg)^2 \sum_{\k\q} \frac{1}{E_{\k\q}^2},
\end{equation}
where we have only considered the leading order term in $a_\mathrm{BF}$, thus demonstrating $\delta n\propto a_\mathrm{BF}^2$. 

Likewise the leading order form of
$\chi_\q$ corresponds to
\begin{equation}
    \chi_\q \simeq  \frac{2\pi a_\mathrm{BF}}{m_r} + 
    \bigg(\frac{2\pi a_\mathrm{BF}}{m_r}\bigg)^2
    \Bigg(
    \sum_{\k\q} \frac1
    {\epsilon_{\q-\k}^\mathrm{B} + \epsilon_{\k}^\mathrm{F}
    -\epsilon_{\q}^\mathrm{F} + g_\mathrm{BB} n_\mathrm{B}}
    -n_\mathrm{F} 
    \sum_\p \frac{1}{\epsilon_\p^\mathrm{F} +\epsilon_\p^\mathrm{B} }
    \Bigg),
\end{equation}
where in the $\mathcal{O}(a_\mathrm{BF}^2)$ term we can take $\mu_\mathrm{B} \simeq g_\mathrm{BB} n_\mathrm{B}$, and neglect $-2g_\mathrm{BB}\delta n$ which is higher order. %

Inspecting the terms in Eq.~\eqref{eq:EdensSM}, we recognize that $\delta n \sum_\q \chi_\q$, $\delta n^2$ and $(2\pi a_\mathrm{BF}/m_r) n_\mathrm{F} \delta n$ are all higher order in $a_\mathrm{BF}$. Therefore, to second order in $a_\mathrm{BF}$ we have
\begin{align} \nn
    \mathcal{E}\simeq&\frac{3}{5} E_\mathrm{F} n_\mathrm{F} 
    + \frac{g_\mathrm{BB}}{2} n_\mathrm{B}^2 
    + \frac{2 \pi a_\mathrm{BF}}{m_\mathrm{r}} n_\mathrm{F} n_\mathrm{B}
    \\
    &- \bigg(  \frac{2\pi a_\mathrm{BF}}{m_\mathrm{r}} \bigg)^2
    n_\mathrm{B} 
    \Bigg(
    \sum_{\k\q} \frac1
    {\epsilon_{\q-\k}^\mathrm{B} + \epsilon_{\k}^\mathrm{F}
    -\epsilon_{\q}^\mathrm{F} + g_\mathrm{BB} n_\mathrm{B}}
    -n_\mathrm{F} 
    \sum_\p \frac{1}{\epsilon_\p^\mathrm{F} +\epsilon_\p^\mathrm{B} }
    \Bigg).
\end{align}

Here, the first line is precisely the mean-field free energy obtained in Ref.~\cite{Viverit2000}, while the second line is the second-order term. 
This expression matches that of Ref.~\cite{Viverit2002} in the limit $n_\mathrm{B}a_\mathrm{BB}^3\ll1$. Importantly, since the free energy reproduces second-order perturbation theory in $a_\mathrm{BF}$, so do the derived quantities $\mu_\mathrm{B}$ and $\mu_\mathrm{F}$.

We emphasize that the second-order perturbation theory of Viverit \emph{et al.}~\cite{Viverit2002} provided the basis for the prediction of Bose-Fermi droplets by Rakshit \emph{et al.} in Ref.~\cite{Rakshit2019SciPost}. In that sense, the perturbative approach employed in Ref.~\cite{Rakshit2019SciPost} is contained within our strong-coupling theory up to lowest order in $n_\mathrm{B}a_\mathrm{BB}^3$. It is therefore of interest to compare the critical values of the various physical parameters at the point of droplet formation, and to check whether the system can reliably be described within perturbation theory.

Table~\ref{tab:criticaltab} shows our calculated values of the critical boson-fermion interaction strength beyond which we predict the existence of droplets. Crucially, we see that across a range of mass ratios, the critical value $k_\mathrm{F}a_\mathrm{BF}$ is close to $-1$, thus highlighting that we are formally beyond the weak-coupling regime. This is reflected in the  densities at the critical point, which we generally find are significantly larger than in Ref.~\cite{Rakshit2019SciPost}: By a factor $7$ for bosons and $10$ for fermions in the case of a K-K mixture, to factors of $32$ and $54$, respectively, for a Cs-Li mixture. On the other hand, we see that the gas parameter $n_\mathrm{B}a_\mathrm{BB}^3$ at the critical interaction strength is minimal for systems with mass ratios of order 1, further supporting our use of a mean-field interaction between bosons.

\begin{table}[ht]
    \centering
    \renewcommand{\arraystretch}{1.3}
    \begin{tabular}{|c||c|c|c|c|c|c|}
\hline
Mixture & $a_{\mathrm{BF}}/a_{\mathrm{BB}}$ & $n_{\mathrm{B}} a_{\mathrm{BB}}^3$
& $n_0/n_\mathrm{B}$
& $n_{\mathrm{F}}/n_{\mathrm{B}}$
& $k_{\mathrm{F}}a_{\mathrm{BF}}$
& $m_{\mathrm{B}}/m_{\mathrm{F}}$ \\
\hline\hline
${}^{23}\mathrm{Na}$-${}^{40}\mathrm{K}$ & $-9.54$ & $6.96\times10^{-5}$ & 0.972 & 0.287 & $-1.01$ & 0.575 \\
${}^{41}\mathrm{K}$-${}^{40}\mathrm{K}$ & $-10.8$ & $6.30\times10^{-5}$ & 0.976 & 0.203 & $-0.982$ & 1.025 \\
${}^{87}\mathrm{Rb}$-${}^{40}\mathrm{K}$ & $-9.60$ & $1.15\times10^{-4}$ & 0.978 & 0.151 & $-0.969$ & 2.175 \\
${}^{133}\mathrm{Cs}$-${}^{6}\mathrm{Li}$ & $-1.84$ & $2.29\times10^{-2}$ & 0.979 & 0.113 & $-0.987$ & 22.095 \\
\hline
    \end{tabular}
    \caption{Values of the various physical parameters at the critical scattering length $a_\mathrm{BF}$ that signals the onset of droplet formation. In the case of the Na-K mixture, this critical point corresponds to the left-most purple circle in Fig.~\ref{fig:mubAbfDroplet}(a) of the main text.}
    \label{tab:criticaltab}
\end{table}

\end{document}